\documentclass[lettersize,journal]{IEEEtran}
\usepackage{amsmath,amsfonts}
\usepackage{array}
\usepackage[caption=false,font=normalsize,labelfont=sf,textfont=sf]{subfig}
\usepackage{textcomp}
\usepackage{stfloats}
\usepackage{url}
\usepackage{verbatim}
\usepackage{graphicx}
\usepackage{cite}
\usepackage{balance}
\usepackage{layout}
\usepackage{subfig}
\usepackage{hyperref}
\usepackage{setspace}
\usepackage{hyperref}
\usepackage{booktabs}
\usepackage{algorithm}    
\usepackage{algpseudocode}     
\usepackage{multirow}
\usepackage{listings, xcolor}
\usepackage{tcolorbox}

\usepackage{url}

\usepackage{ragged2e}

\definecolor{verylightgray}{rgb}{.97,.97,.97}

\lstdefinelanguage{Solidity}{
	keywords=[1]{anonymous, assembly, assert, balance, break, call, callcode, case, catch, class, constant, continue, constructor, contract, debugger, default, delegatecall, delete, do, else, emit, event, experimental, export, external, false, finally, for, function, gas, if, implements, import, in, indexed, instanceof, interface, internal, is, length, library, log0, log1, log2, log3, log4, memory, modifier, new, payable, pragma, private, protected, public, pure, push, require, return, returns, revert, selfdestruct, send, solidity, storage, struct, suicide, super, switch, then, this, throw, transfer, true, try, typeof, using, value, view, while, with, addmod, ecrecover, keccak256, mulmod, ripemd160, sha256, sha3}, 
	keywordstyle=[1]\color{blue}\bfseries,
	keywords=[2]{address, bool, byte, bytes, bytes1, bytes2, bytes3, bytes4, bytes5, bytes6, bytes7, bytes8, bytes9, bytes10, bytes11, bytes12, bytes13, bytes14, bytes15, bytes16, bytes17, bytes18, bytes19, bytes20, bytes21, bytes22, bytes23, bytes24, bytes25, bytes26, bytes27, bytes28, bytes29, bytes30, bytes31, bytes32, enum, int, int8, int16, int24, int32, int40, int48, int56, int64, int72, int80, int88, int96, int104, int112, int120, int128, int136, int144, int152, int160, int168, int176, int184, int192, int200, int208, int216, int224, int232, int240, int248, int256, mapping, string, uint, uint8, uint16, uint24, uint32, uint40, uint48, uint56, uint64, uint72, uint80, uint88, uint96, uint104, uint112, uint120, uint128, uint136, uint144, uint152, uint160, uint168, uint176, uint184, uint192, uint200, uint208, uint216, uint224, uint232, uint240, uint248, uint256, var, void, ether, finney, szabo, wei, days, hours, minutes, seconds, weeks, years},	
	keywordstyle=[2]\color{teal}\bfseries,
	keywords=[3]{block, blockhash, coinbase, difficulty, gaslimit, number, timestamp, msg, data, gas, sender, sig, value, now, tx, gasprice, origin},	
	keywordstyle=[3]\color{violet}\bfseries,
	identifierstyle=\color{black},
	sensitive=true,
	comment=[l]{//},
	morecomment=[s]{/*}{*/},
	commentstyle=\color{gray}\ttfamily,
	stringstyle=\color{red}\ttfamily,
	morestring=[b]',
	morestring=[b]"
}

\begin{document}

\title{Beyond the Hype: A Large-Scale Empirical Analysis of On-Chain Transactions in NFT Scams}%

\author{Wenkai~Li$^{*}$,
        Zongwei~Li$^{*}$,
        Xiaoqi~Li,
        Chunyi~Zhang,
        Xiaoyan~Zhang,
        and~Yuqing~Zhang
%
\IEEEcompsocitemizethanks{%
\IEEEcompsocthanksitem Manuscript received xx xxxx. This work is sponsored by the National Natural Science Foundation of China (No.62362021 and No.62402146), CCF-Tencent Rhino-Bird Open Research Fund (No.RAGR20230115), and Hainan Provincial Department of Education Project (No.HNJG2023-10). \textit{(Corresponding author: Xiaoqi Li.)}
\IEEEcompsocthanksitem Wenkai Li, Zongwei Li, Xiaoqi Li, Chunyi Zhang, and Xiaoyan Zhang are with the School of Cyberspace Security, Hainan University, Haikou, 570228, China (e-mail: cswkli@hainanu.edu.cn; lizw1017@hainanu.edu.cn; csxqli@ieee.org; zhangchunyi@hainanu.edu.cn; xia0yanZhang@hainanu.edu.cn).
\IEEEcompsocthanksitem Yuqing Zhang is with the University of Chinese Academy of Sciences, Beijing, 100049, China (e-mail: zhangyq@nipc.org.cn).

}
%
\thanks{$^{*}$These authors contributed equally to this work.}
}
\markboth{Journal of \LaTeX\ Class Files,~Vol.~14, No.~8, August~2021}%
{Shell \MakeLowercase{\textit{et al.}}: A Sample Article Using IEEEtran.cls for IEEE Journals}


\maketitle

\begin{abstract}
Non-fungible tokens (NFTs) serve as a representative form of digital asset ownership and have attracted numerous investors, creators, and tech enthusiasts in recent years. However, related fraud activities, especially phishing scams, have caused significant property losses. There are many graph analysis methods to detect malicious scam incidents, but no research on the transaction patterns of the NFT scams. Therefore, to fill this gap, we are the first to systematically explore NFT phishing frauds through graph analysis, aiming to comprehensively investigate the characteristics and patterns of NFT phishing frauds on the transaction graph. During the research process, we collect transaction records, log data, and security reports related to NFT phishing incidents published on multiple platforms. After collecting, sanitizing, and unifying the data, we construct a transaction graph and analyze the distribution, transaction features, and interaction patterns of NFT phishing scams. We find that normal transactions on the blockchain accounted for 96.71\% of all transactions. Although phishing-related accounts accounted for only 0.94\% of the total accounts, they appeared in 8.36\% of the transaction scenarios, and their interaction probability with normal accounts is significantly higher in large-scale transaction networks. Moreover, NFT phishing scammers often carry out fraud in a collective manner, targeting specific accounts, tend to interact with victims through multiple token standards, have shorter transaction cycles than normal transactions, and involve more multi-party transactions. This study reveals the core behavioral features of NFT phishing scams, providing important references for the detection and prevention of NFT phishing scams in the future.
\end{abstract}


\section{Introduction}

\IEEEPARstart{N}{on-fungible} tokens (NFTs)~\cite{jones2021scientists} have emerged as a transformative representation of digital ownership since 2021~\cite{kostick2022nfts, upadhyay2025dark}. The projected market size for NFTs is expected to exceed 26.41 billion USD by 2024~\cite{NFTMarket2025}, highlighting their capacity to attract substantial engagement from a diverse array of participants, including global investors, artists, and technology enthusiasts~\cite{upadhyay2025dark}. The financial impact of fraudulent activities within the NFT ecosystem is substantial. A report from Elliptic revealed that between July 2021 and July 2022, illicit gains from NFT scams exceeded \$100 million, with an average of approximately \$300,000 per incident~\cite{NFTsFinancialCrime2025}. This trend has not only persisted but has also evolved in its operational tactics. More importantly, in 2023, there is a significant development where threat actors began enlisting minors to execute large-scale phishing attacks~\cite{roy2024unveiling, cheon2025scarcity}. These operations led to the compromise of over 32,000 wallets and 900 Discord servers, resulting in the theft of assets valued at \$73 million~\cite{PhishingFrenzy2025}.

\begin{figure}[!t]
    \centering
    \includegraphics[width=0.5\textwidth]{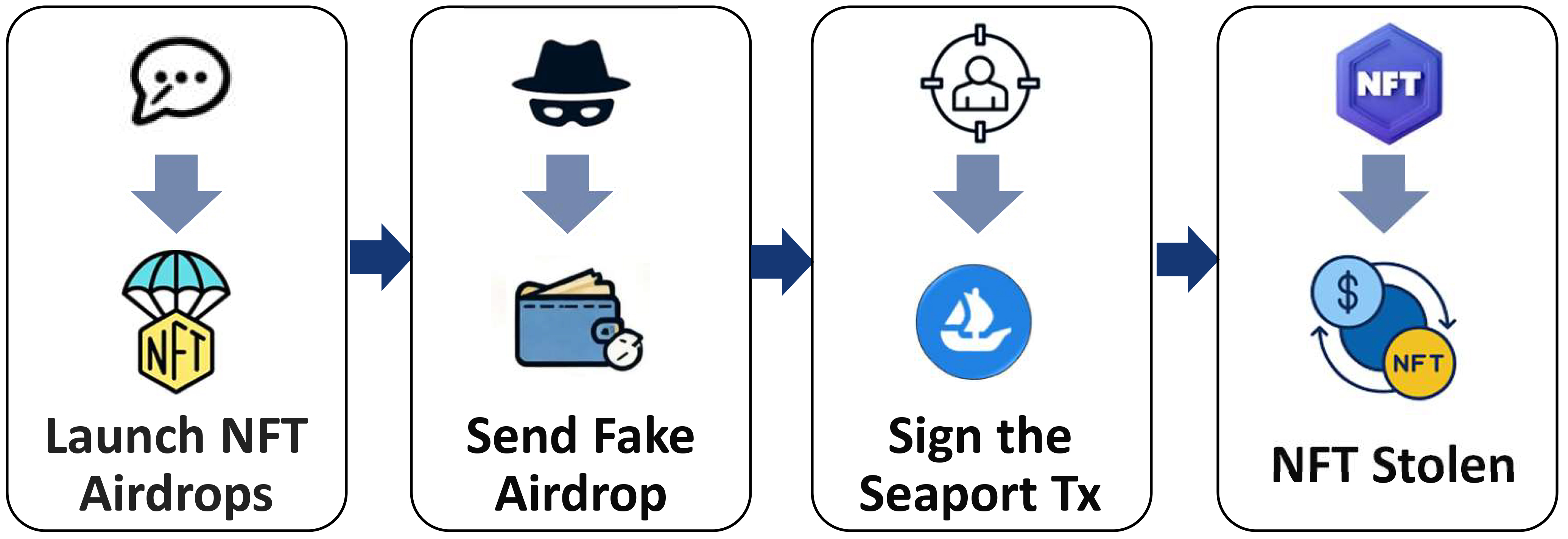}
    \caption{The Motivation Case of the NFT Airdrop Phishing.}
    \vspace{-2ex}
    \label{fig:motivation_case}
\end{figure}

The rapid evolution of blockchain technology has transformed NFTs from simple digital collectibles~\cite{sung2023nft} into a diverse ecosystem~\cite{alkhader2023leveraging,guidi2023nft} encompassing art, gaming, finance, and intellectual property. By leveraging the inherent transparency of distributed ledgers, NFTs provide verifiable proof of ownership and enhance transaction integrity. Consequently, trading platforms have facilitated substantial transaction volumes~\cite{tang2023exploring}, attracting widespread investment and catalyzing the transition from traditional supply chains to metaverse economies. Despite the transparency inherent in distributed ledgers, the ecosystem remains susceptible to exploitation due to user anonymity and intricate smart contract interactions. This environment has fostered a proliferation of fraudulent activities~\cite{zhang2025security, li2025penetrating, zhang2025penetration}, such as rug pulls~\cite{zhou2024stop}, phishing~\cite{yang2024stole}, and wash trading~\cite{von2022nft}, posing severe security challenges to the NFT landscape. For instance, as Figure \ref{fig:motivation_case} shows, at the same time as the announcement of promotional NFT airdrops~\cite{allen2023airdrop}, some malicious fraudsters executed a targeted phishing campaign~\cite{NFTAirdropPhishing2025}. By exploiting the Seaport protocol~\cite{Seaport2025} to bypass the security detection mechanism, the attackers distributed bait to targeted wallets. Victims were ultimately directed to phishing websites, where signing a deceptive confirmation instruction resulted in the unauthorized transfer of their assets~\cite{zhou2024artemis}.

The foundational studies primarily centered on the economic feasibility~\cite{sadorsky2024time,kong2024characterizing}, asset valuation~\cite{kapoor2022tweetboost, fekih2023formal}, and statistical properties of NFTs~\cite{qian2022understanding}. However, as the ecosystem has matured, the maturing market has exposed critical security gaps. Recent studies have thus shifted their focus to technical and behavioral threats. For example, some researchers \cite{ma2025uncovering,xiao2025wakemint,yang2023definition,niu2025natlm} systematically classified vulnerabilities from the code aspect~\cite{wu2025security, peng2025mining}, and then identified compromised contracts~\cite{li2025interaction, luo2025movescanner}. Recent advancements also leverage AI and LLMs for security analysis~\cite{wang2025ai, peng2025multicfv, gong2025information, zhang2025risk, xiang2025security}. Additionally, some studies~\cite{chen2024dark,huang2024unveiling, wen2023nftdisk} examined behavioral fraud, exploring various forms of wash trading activities within the NFT marketplace.

Despite the significant contributions made by existing literature, there remains a notable lack of systematic inquiry into the on-chain transaction dynamics associated with NFT phishing scams. Thus, we aim to conduct a comprehensive analysis of NFT phishing behaviors utilizing a large-scale dataset of on-chain transactions. Our investigation is structured around four central research questions (RQs):

\begin{itemize}
\item[RQ1] \textbf{(Statistics):} How is the distribution of NFT phishing? 
\item[RQ2] \textbf{(Transaction Features): } What are the transaction features of NFT phishing under different ERC token standard protocols?
\item[RQ3] \textbf{(Interaction Patterns):} How does NFT phishing behave?


\end{itemize}

To support our study, we adopt the data collection method described by \cite{yang2024stole} and conduct a thorough analysis of various indicators associated with NFT phishing. Initially, we gather transaction and log data from NFT phishing cases and security reports published on several platforms, including Chainabuse~\cite{Chainabuse2025}, Twitter/X~\cite{Twitter2025}, SlowMist~\cite{SlowMist2025}, PeckShield~\cite{Peckshield2025}, and PANews~\cite{Panewslab2025}. Furthermore, we acquire transaction data for the phishing addresses listed on ScamSniffer~\cite{ScamSniffer2025}, Etherscan~\cite{Etherscan2025}, and Chainabuse~\cite{Chainabuse2025}, focusing on the first 18 million blocks of the Ethereum network. After collecting the data, we conduct Data sanitization and unification (i.e., data loading, field alignment, structure alignment, and type conversion), remove the data with incomplete fields, and standardize different types of transactions (i.e., ERC20, ERC721, ERC1155). To enhance our understanding of NFT phishing within the transaction graph framework, we conduct an empirical investigation of the existing transaction data and associated events. The primary objective is to address four key research questions. Our analysis indicates that normal transactions accounted for 96.71\% of total transactions, suggesting a predominance of legitimate activity. Furthermore, our findings reveal that phishers frequently engaged in collusive practices, operating in groups to perpetrate fraudulent activities. Moreover, we find that phishers tend to interact with the same victim through multiple ERC protocols to spread the risk of being traced.

It is essential to underscore that our research is centered on elucidating the phenomena and patterns associated with NFT phishing as observed in the transaction graphs. We entail a systematic and comprehensive analysis of the data patterns inherent within the graph, coupled with empirical investigations into NFT phishing scams. Based on the insights derived from our analytical assessments, we assert that we have significantly enhanced the interpretability of NFT phishing within the transaction graph framework. This advancement holds the potential to direct future research initiatives aimed at the detection of NFT phishing scams through transaction graph-based methodologies.

Our contributions are as follows:
\begin{itemize}
\item To the best of our knowledge, this study represents the first empirical analysis of the NFT phishing scam graph. Our research provides substantial evidence that NFT phishing scams can be effectively analyzed through the lens of graph theory.

\item Our findings reveal that although the incidence of NFT phishing attacks on the blockchain is high, a significant 96.71\% of transaction environments remain secure. NFT phishers frequently engage in coordinated actions aimed at specific account targets.

\item We observe that NFT fraudsters are inclined to execute multi-protocol mixed transactions with their victims, and the duration of the phishing transaction cycle is shorter than that of standard transactions. Moreover, NFT scammers are involved in a greater number of multi-party transactions than typical accounts.

\end{itemize}

The remainder of the article is organized as follows: The Section \ref{sec:background} offers key foundational insights into the blockchain ecosystem and NFT transactions. The Section \ref{sec:related_work} reviews relevant literature. In Section \ref{sec:data_handle}, we present a detailed procedure for data processing and standardization. Section \ref{sec:findings} analyzes the primary characteristics of NFT phishing scams by addressing four critical questions. Finally, in Section \ref{sec:conclusion}, we provide a conclusion and explore future directions.

\section{Background}
\label{sec:background}

In this section, we introduce some basic background knowledge of the blockchain ecosystem and NFT transactions.

\subsection{The Blockchain Ecosystem}
Blockchain technology serves as a decentralized, peer-to-peer network system that forms the backbone of the modern Web3 ecosystem~\cite{toufaily2024blockchain}. By employing advanced cryptographic techniques and distributed consensus mechanisms, blockchain ensures the integrity, transparency, and immutability of on-chain data~\cite{he2023tokenaware}, eliminating the need for a central authority~\cite{sun2024doubleup}. Built upon this secure framework, smart contracts operate as autonomous, Turing-complete programs that execute predefined logic upon the fulfillment of specific conditions~\cite{cai2025detecting}. This programmability has spurred the growth of Decentralized Applications (DApps)~\cite{zheng2025dappcheat}, which are now seamlessly integrated across various industrial sectors. Therefore, a diverse multi-chain ecosystem used for creating and trading NFTs has emerged~\cite{li2024famulet}, supported by leading platforms such as Ethereum~\cite{gao2025implementation}, Solana~\cite{shen2025blockchain}, Cardano, EOS~\cite{liu2025empirical}, Tezos, and Polygon~\cite{huang2025comparative}.

\subsection{The NFT Transactions}
NFT (Non-Fungible Token) is a technology based on blockchain that realizes the unique ownership of digital assets~\cite{ma2025uncovering}. It provides proof of ownership for digital assets such as digital artworks, collectibles, and game coins through a unique identifier on the blockchain~\cite{song2025automated}. NFTs are often used to represent digital products such as digital artworks, collectibles, and game coins~\cite{yang2023definition}. The use of blockchain records data such as ownership changes, transaction history, and minting information of NFTs to ensure the transparency and immutability of the data. From a technical perspective, the transactional nature of NFTs relies on standardized smart contracts~\cite{liang2024identity}. The mainstream solution is based on the Ethereum improvement proposal EIP (e.g., ERC721, ERC1155) \cite{tan2024bubble}, which is adapted to different asset management methods. The ERC721 standard~\cite{tan2024bubble} is the most popular NFT contract standard and is often used for single rare NFTs such as artworks, virtual land, and collectibles (e.g., BAYC~\cite{lee2024exploring}, Decentraland~\cite{nnamonu2023digital}). Listing \ref{lst:NFT_transfer} shows the functions related to NFT transfers. Each \textit{tokenId} and the contract address are jointly used to represent a unique NFT on the blockchain. It includes interfaces such as \textit{balanceOf}, \textit{ownerOf}, and \textit{safeTransferFrom} for querying balance, querying owner, and secure transfer. The ERC1155 standard's~\cite{tan2024bubble} efficient and low-cost mechanism is suitable for large-scale applications in hybrid asset systems, which also support NFTs and fungible tokens. It is commonly used for game assets, batch-issued collections, etc. NFTs (e.g., Axie Infinity~\cite{ali2025playing}, NBA Top Shot~\cite{lee2022measuring}). It includes the \textit{safeBatchTransferFrom} interface to support batch transfers, saving gas fees. 

\begin{lstlisting}[language=Solidity, 
                    basicstyle=\fontsize{8}{9}\ttfamily,
                    numbers=left,
                    captionpos=b,
                    commentstyle=\color{orange},
                    aboveskip = 1em,
                    belowskip = 1em,
                    numbersep= -1em,
                    caption= NFT transfer-related functions standardized by ERC721,
                    label=lst:NFT_transfer,
                    ] 
  function safeTransferFrom(address _from, address _to, uint256 _tokenId, bytes data) external payable;
  function safeTransferFrom(address _from, address _to, uint256 _tokenId) external payable;
  function transferFrom(address _from, address _to, uint256 _tokenId) external payable;
  function approve(address _approved, uint256 _tokenId) external payable;
\end{lstlisting}

One defining feature of the NFT ecosystem is its cross-platform interoperability~\cite{belchior2021survey}, supported by universal token standards such as ERC-721~\cite{tan2024bubble}. This standardization empowers users with complete control over their digital assets, facilitating permissionless portability across diverse marketplaces without dependence on centralized authorization~\cite{peng2021privacy}. For instance, an in-game asset acquired on OpenSea~\cite{white2022characterizing} can be effortlessly listed and resold on a rival platform like LooksRare~\cite{huang2024unveiling}. In terms of transaction execution, secondary markets primarily utilize fixed-price listings and auction mechanisms~\cite{yang2024stole}. In a fixed-price model, the transaction occurs in a single step, with the buyer transferring the total amount, including the asset price and platform fees, to obtain immediate ownership. Conversely, auctions operate under an ascending bid structure, where the asset is awarded to the highest bidder. In an auction scenario, buyers submit incremental bids, and the transaction concludes only when the seller accepts the highest offer, at which point platform fees are levied on the seller's revenue.

\section{Related Work}
\label{sec:related_work}

\subsection{Blockchain Transaction Graph Analysis}
To understand the behaviors of various entities in blockchains, many previous studies~\cite{Chen2018Understanding,qi2023blockchain,luo2024multi,wu2023tracer,geng2022novel,guo2023graph,du2023breaking,wu2023know,wu2023understanding,wu2021detecting} have utilized graph and motif structures to interpret the transaction data. Chen et al.~\cite{Chen2018Understanding} are the first to use a visual mapping to analyze the fund flows, contract creation, and contract interactions within Ethereum, thereby revealing the behavioral characteristics and interaction patterns of users. This approach relies on graph analysis technology to convert high-dimensional complex relationships into intuitive graphs, thereby reducing the difficulty of network behavior analysis. Wu et al.~\cite{wu2023tracer} transforms blockchain transaction data into a weighted directed graph of fund transfers and asset exchanges. It enhances the account traceability capability by dynamically evaluating the relationships between associated nodes and combines with the local community detection algorithm to identify money laundering clusters. This research analyzed money laundering transactions and promoted the development of tracking technologies for money laundering behaviors. Wu et al.~\cite{wu2023know} proposed the MoTS method, which constructed a universal transaction semantic representation method for the Web3 blockchain ecosystem, enabling the modeling of transaction relationships such as internal transactions, external transactions, ERC20/721/1155 transactions, etc. By generating vectors based on the frequency characteristics of 16 types of motifs, various semantic meanings such as transfer, swap, and mint can be recognized. This research offers novel insights into methods such as understanding blockchain transactions and detecting fraudulent accounts. Wu et al.~\cite{wu2023understanding} were pioneers in the analysis of the structural dynamics of the self-trading network associated with blockchain accounts, investigating the behavioral evolution across the account lifecycle. Their research meticulously examines both the dynamic behaviors, such as lifecycle progression, transaction frequency, monetary flow, and patterns of neighboring interactions. And the static micro-characteristics, including transaction network density, local clustering coefficients, and interaction patterns. By distinguishing between legitimate and illicit accounts, this study introduces a novel technological framework for blockchain account detection. By initiating the analysis with the self-network of the account, their work provides enriched insights into the behavioral patterns exhibited by blockchain accounts. 

\subsection{NFT Scam Analysis}
The research subjects of NFT scams mainly include rug pulls~\cite{sharma2023understanding, huang2023miracle,sharma2025discovering,roy2024unveiling}, wash trading~\cite{ chen2024dark,niu2024unveiling,wen2023nftdisk,huang2024unveiling}, and others~\cite{roy2024unveiling,huang2024unveiling,yang2024stole}. 

Sharma et al.~\cite{sharma2025discovering} construct 8 typical fraudulent transaction patterns related to rug pull, and used the Graph Isomorphism Networks (GIN) model to detect rug pull. Roy et al.~\cite{roy2024unveiling} are the first to conduct a study on fraudulent projects in NFT promotion on Twitter. It is discovered that fraudsters used robot accounts to artificially inflate the popularity of the projects, thereby luring users into making investments. The current anti-fraud measures have failed to address the issues of low coverage and slow response speed in such promotion scams, providing a systematic analytical basis for the study of NFT promotion scams. Chen et al.~\cite{chen2024dark} are the first to explore three trading patterns (i.e., Round-trip Trading, Unprofitable Trading, and Hidden Trading), and they also design corresponding detection algorithms. After conducting a quantitative analysis, it identifies the three main reasons for the significant losses caused by NFT wash trading, including the incentive mechanisms of platforms (e.g., LooksRare) that encourage users to engage in false transactions, the design flaws of NFTs, and the dominance of a small number of accounts in the majority of wash trading. This study quantitatively analyzes the causes and corresponding losses of NFT wash trading, providing targeted governance suggestions that offer a direction for future research. Yang et al.~\cite{yang2024stole} construct an NFT phishing fraud dataset, classifying it into four typical patterns (i.e., NFT atmosphere fraud signatures, fraud authorizations, identity credential theft, and induced transfers). It has been found that NFT fraudsters tend to target high-value NFT collections first, similar to the Ethereum Domain Name Service (ENS), etc. Meanwhile, the study combines the transaction features and NFT features with a random forest for detection. It analyzes the importance of various feature types in the detection process, providing a technical reference for NFT phishing detection.

Different from the above studies, we are the first to investigate various characteristics of NFT phishing scams on the transaction graph.

\section{Data Sanitization and Unification}
\label{sec:data_handle}
In this section, we mainly introduce how we process the data and construct the transaction graph structure. 

Firstly, we collect NFT Phishing scams following \cite{yang2024stole}. 
We systematically collect transaction and log data about NFT phishing incidents, along with comprehensive security reports published on various platforms, including Chainabuse~\cite{Chainabuse2025}, Twitter/X~\cite{Twitter2025}, SlowMist~\cite{SlowMist2025}, PeckShield~\cite{Peckshield2025}, and PANews~\cite{Panewslab2025}. Furthermore, we obtain transaction data from phishing addresses identified on ScamSniffer~\cite{ScamSniffer2025}, Etherscan~\cite{Etherscan2025}, and Chainabuse~\cite{Chainabuse2025}, with a particular emphasis on the first 18 million blocks of the Ethereum network via the BlockchainSpider~\cite{wu2023tracer}. This focused approach enables us to analyze and discern patterns in phishing activities effectively.

The data of the three types of token standards (ERC20, ERC721, ERC1155) uniformly follow the same four-stage processing framework, as shown in the algorithm \ref{alg:unify}, including data loading, field Alignment, structure alignment, and type conversion.

In Stage 1 and Stage 2, normal account data and phishing account data are loaded, respectively. Using \textit{dtype=str} to avoid type inference errors, the \textit{label} and \textit{source\_file} fields are added, and the field renaming mapping is applied (such as \textit{blocknum} $\rightarrow$ \textit{blockNum}, \textit{nft\_from} $\rightarrow$ \textit{from}). In Stage 3, field alignment is performed, a unified field set $\mathcal{C}$ is defined, missing fields are filled with \textit{None}, and rearranged in a unified order. In Stage 4, the data is vertically concatenated and type conversions are carried out: \textit{blockNum} is converted to an integer, the \textit{timestamp} of NFT data is converted to a datetime, and numerical fields are converted to the corresponding numerical types.

\subsection{Transaction Graph Construction}
Firstly, we load the multi-protocol transaction data based on a unified format CSV file, focusing on the key fields of the graph ( i.e., transaction \textit{hash}, sender \textit{from}, receiver \textit{to}, block number \textit{blockNum}, transaction amount \textit{value}, token contract address \textit{token\_contract}, NFT series \textit{series}, etc.) to optimize the efficiency of memory usage. For the differences in fields among different protocols, we achieve data format uniformity through a standardized process and annotate the corresponding protocol type for each transaction record. 

\begin{algorithm}[H]
\caption{Token Transaction Data Unification}
\label{alg:unify}
\begin{algorithmic}[1]
\Require \\Normal accounts $D_{\text{normal}}$, \\
Phishing accounts $D_{\text{phishing}}$,\\
Token Types $T \in \{\text{ERC20}, \text{ERC721}, \text{ERC1155}\}$
\Ensure Unified data $\mathcal{D}_{\text{unified}}$

\State \textbf{Stage 1: Loading Normal Accounts}
\State $\mathcal{F}_{\text{normal}} \gets$ match all $T$ types in $D_{\text{normal}}$
\For{each $f \in \mathcal{F}_{\text{normal}}$}
    \State $df \gets$ read $f$ as String
    \State $df[\text{label}] \gets \text{"normal"}$
    \State $df[\text{source\_file}] \gets f.\text{name}$
    \State Field renaming mapping $\mathcal{M}_{\text{normal}}^T$
    \State Add $df$ to $\mathcal{D}_{\text{normal}}$
\EndFor
\State $D_{\text{normal}} \gets$ merge $\mathcal{D}_{\text{normal}}$ 

\State \textbf{Stage 2: Loading Phishing Accounts}
\State $\mathcal{F}_{\text{phishing}} \gets$ match all $T$ types in $D_{\text{phishing}}$
\For{each $f \in \mathcal{F}_{\text{phishing}}$}
    \State $df \gets$ read $f$ as String
    \State $df[\text{label}] \gets \text{"phishing"}$
    \State $df[\text{source\_file}] \gets f.\text{name}$
    \State Field renaming mapping $\mathcal{M}_{\text{phishing}}^T$
    \State Add $df$ to $\mathcal{D}_{\text{phishing}}$
\EndFor
\State $D_{\text{phishing}} \gets$ merge  $\mathcal{D}_{\text{phishing}}$ 

\State \textbf{Stage 3: Field Alignment}
\State $\mathcal{C} \gets$ define unified field set (relying on $T$)
\For{each $D \in \{D_{\text{normal}}, D_{\text{phishing}}\}$}
    \For{each $c \in \mathcal{C}$}
        \If{$c \notin D.\text{columns}$}
            \State $D[c] \gets \text{None}$
        \EndIf
    \EndFor
    \State $D \gets D[\mathcal{C}]$ \Comment{Reorder in the unified sequence}
\EndFor

\State \textbf{Stage 4: Merge and Type Conversion}
\State $\mathcal{D}_{\text{unified}} \gets$ vertical joining $D_{\text{normal}}$ and $D_{\text{phishing}}$
\State $\mathcal{D}_{\text{unified}}[\text{blockNum}] \gets$ apply \textit{hex\_to\_int}
\If{$T \in \{\text{ERC721}, \text{ERC1155}\}$}
    \State $\mathcal{D}_{\text{unified}}[\text{timestamp}] \gets$ apply \textit{hex\_to\_datetime}
    \State $\mathcal{D}_{\text{unified}}[\text{nft\_num}] \gets$ convert to numeric type
\EndIf
\If{$T = \text{ERC20}$}
    \State $\mathcal{D}_{\text{unified}}[\text{value}] \gets$ convert to numeric type
\EndIf

\State \Return $\mathcal{D}_{\text{unified}}$
\end{algorithmic}
\end{algorithm}

Then, to avoid the influence of large connected components on the analytical focus, a dual filtering strategy is implemented. We eliminate system-level addresses such as zero addresses and dead addresses. Meanwhile, we filter super hub nodes with a degree exceeding the threshold, which is set to 30 by default. 

Finally, for each pair of addresses, multiple transactions are aggregated according to the protocol type, merging duplicate transactions into a single edge. And then extracting multi-dimensional edge features, including transaction count \textit{tx\_count}, total transaction volume \textit{total\_amount}, average transaction size \textit{avg\_amount}, transaction time span \textit{block\_span}, number of involved contracts and series shown in Table \ref {tab:edge_features}. Through an outer join to fuse the edge feature tables of the three types of protocols, a complete edge attribute set containing exclusive features and cross-protocol statistical information is formed.

\begin{table}[htbp]
\centering
\caption{The Description of Edge Features}
\label{tab:edge_features}
\resizebox{\columnwidth}{!}{
\begin{tabular}{ll}
\toprule
\textbf{Feature Names} & \textbf{Description} \\
\midrule
tx\_count & the number of transactions \\
total\_amount & total transactions value/NFT amount \\
avg\_amount & average transaction value/NFT amount \\
first\_block & the first transaction block number \\
last\_block & the last transaction block number \\
block\_span & last\_block minus first\_block \\
token\_contract\_count & involved number of token contracts  \\
series\_count & involved number of NFT series  \\
\bottomrule
\end{tabular}
}

\end{table}

We employ an undirected graph modeling strategy, which ignores the flow of funds and focuses on the transaction associations between addresses, building a transaction graph based on the cuDF~\cite{cuDF2025} and cuGraph library~\cite{cugraph2025}. We regard all accounts as nodes and all transactions as edges. The weakly connected components algorithm is used to extract all the sets of interrelated addresses. 
Based on the characteristics of the node labels within the connected components, the graphs are classified into the following three categories. (1) Phishing graph that contains only phishing addresses; (2) Normal graph that contains only normal addresses; (3) Mixed graph that contains both phishing addresses and normal addresses.

Set a minimum node count threshold (default is 10), eliminate isolated or overly small connected components, and retain large graphs with a statistical analysis value for subsequent in-depth exploration.

\section{Findings}
\label{sec:findings}

\subsection{How is the distribution of NFT phishing?}
\label{subsec:Statistics}
After the processing in Section \ref{sec:data_handle}, we have collected 37 NFT phishing accounts and 3898 normal accounts. As shown in the statistical information in Table \ref{tab:graph_overview}, we present the overall statistics and node composition of graphs with a node count of at least 10. This graph has a total of 3,925 weakly connected elements, among which 275 meet the standard of containing at least 10 nodes. There are a total of 3,935 nodes in these 275 graphs.

\begin{table}[htbp]
\centering
\caption{Statistical Information of Graphs, where node count $\geq$ 10}
\label{tab:graph_overview}
\begin{tabular}{lcc}
\toprule
\textbf{Indicators} & \textbf{Amount}  \\
\midrule
 Connected Graphs Amount & 3925  \\
Graph Amount & 275  \\
Graph Nodes & 3935  \\
\midrule
Normal Graphs & 246 \\
Mixed Graph  & 29  \\
\midrule
Normal Accounts & 3898  \\
NFT Phishing Accounts & 37  \\
\bottomrule
\end{tabular}
\vspace{-1em}
\end{table}

Among the graphs, only 246 normal graphs contain ordinary addresses, while 29 mixed graphs contain both ordinary addresses and phisher addresses. Regarding node composition, the large graph includes 3,898 normal addresses and 37 phishing addresses.
Although the number of phishing nodes in graphs is 37, 78.37\% of them exist in the mixed graphs, which indicates that there may be interaction between units and ordinary users. Meanwhile, the 89.45\% graphs only transfer to normal addresses, indicating that the blockchain network is mainly composed of legitimate activities, while phishing activities are relatively concentrated in a specific and interrelated structure.

Table \ref{tab:size_type_comparison} presents the structural attributes differences of different-sized graphs from the cross-dimensional perspective of graph scale (small graph, large graph) and graph type (mixed graph, normal graph).

\begin{table}[htbp]
\centering
\caption{The Comparison of Small and Large Graphs}
\label{tab:size_type_comparison}
\begin{tabular}{lccc}
\toprule
\textbf{Categories} & \textbf{Graph Type} & \textbf{Amount} & \textbf{Ratio} \\
\midrule
\multirow{2}{*}{Small Graph} & mixed\_graph & 100 & 2.74\% \\
 & normal\_graph & 3550 & 97.26\% \\
\hline
\multirow{2}{*}{Large Graph} & mixed\_graph & 29 & 10.55\% \\
 & normal\_graph & 246 & 89.45\% \\
\bottomrule
\end{tabular}
\vspace{-1em}
\end{table}

For a small graph, the number of mixed graphs (graphs containing both normal addresses and phishing addresses) is 100, accounting for only 2.74\%; normal graphs (graphs containing only normal addresses) reach 3550, accounting for 97.26\%. This indicates that in the structure of a small graph, the independent aggregation of normal addresses is the dominant mode, and the cross-group penetration behavior of phishing entities is relatively rare.
In a large graph, the number of mixed graphs is 29, and the proportion increases to 10.55\%; normal graphs are 246, accounting for 89.45\%. Compared with small graphs, the proportion of mixed graphs in large graphs has significantly increased, indicating that phishing entities are more inclined to form interactions with normal addresses in large-scale connected structures. Such graphs have crucial value for victim traceability and analysis of phishing transmission paths.
In summary, from a small graph to a large graph, the proportion of mixed graphs has increased from 2.74\% to 10.55\%, reflecting the scale penetration characteristic of phishing entities in larger graph structures, where the probability of interaction between phishing addresses and normal addresses is higher.

\begin{tcolorbox}[boxrule=1pt,boxsep=1pt,left=3pt,right=3pt,top=3pt,bottom=3pt]
\textbf{Answer to RQ1.}
NFT phishing accounts only account for 0.94\% of the total accounts, but their transactions occur in 8.36\% of the transaction scenarios, and their interaction probability with normal addresses increases 2.85 times in large-scale transaction graphs.
\end{tcolorbox}

\subsection{What are the transaction features of NFT phishing under different ERC token standard protocols?}
\label{subsec:Transaction_Features}
 
To reveal the trading tendencies of NFT phishers in different transaction protocols, we conduct an in-depth analysis of NFT phishing transactions under different ERC standards.

Firstly, we count the number of transactions in the normal transaction graph and those in the transaction graph with phishing accounts (i.e., the mixed graph). As the statistics results in Table \ref{tab:protocol_usage}, in the case of ERC20 tokens, the proportion of edges in the normal transaction graph is 54.30\%, and in the mixed transaction graph, the proportion of edges is 76.34\%. Both of these indicators are significantly higher than those observed in the ERC721 and ERC1155 cases. Moreover, only the proportion of ERC20 transactions exceeded 50\%. This indicates that when conducting ERC721 and ERC1155 transactions, ERC20 transactions are usually conducted simultaneously. In ERC721, the edge ratios of the normal and mixed transaction graphs are not significantly different, but the average value of 1.48 in the mixed transactions is slightly higher than 1.19 in the normal situation. This indicates that in ERC20 and ERC1155, normal transactions are dominant, while ERC721 has more single transactions and higher NFT quantities.

\begin{table}[htbp]
\centering
\caption{Usage of ERC Protocols in Graph Types}
\label{tab:protocol_usage}
\resizebox{\columnwidth}{!}{
\begin{tabular}{llcccc}
\toprule
\textbf{ERCs} & \textbf{Types} & \textbf{Edges} & \textbf{Ratio} & \textbf{Amount}& \textbf{Average} \\
\midrule
\multirow{2}{*}{ERC20} & normal & 1749 & 54.30\% & 7.66e+15 & 4.38e+12 \\
 & mixed & 497 & 76.34\% & 8.30e+07 & 1.67e+05 \\
\midrule
\multirow{2}{*}{ERC721} & normal & 1138 & 35.33\% & 1352.0 & 1.19 \\
 & mixed & 216 & 33.18\% & 320.0 & 1.48 \\
\midrule
\multirow{2}{*}{ERC1155} & normal & 554 & 17.20\% & 596.0 & 1.08 \\
 & mixed & 49 & 7.53\% & 58.0 & 1.21 \\
\bottomrule
\end{tabular}
}
\end{table}

Furthermore, to investigate whether NFT scammers utilize multiple protocols for complex transfers, we summarize the usage rates of different protocols in the transaction graph in Table \ref{tab:cross_protocol_tx}. The cross-protocol transaction ratio of 3.38\% in the mixed graph is higher than the ratio of 2.27\% in the normal graph, indicating that the phishing accounts are more likely to interact with the same address across multiple protocols.

\begin{table}[htbp]
  \centering
  \caption{Cross-protocol Transaction Statistics}
  \label{tab:cross_protocol_tx}
  \resizebox{\columnwidth}{!}{
  \begin{tabular}{lcc}
    \toprule
    \textbf{Graph Types} & \textbf{Protocol Combination} & \textbf{Amount (Ratio)} \\
    \midrule
    \multirow{5}{*}{normal} & ERC20+ERC721    & 58 (1.80\%)  \\
                            & ERC20+ERC1155   & 10 (0.31\%)  \\
                            & ERC721+ERC1155  & 15 (0.47\%)  \\
                            & at least two protocols      & 73 (2.27\%)  \\
                            & three protocols         & 5 (0.16\%)   \\
    \midrule
    \multirow{5}{*}{mixed}  & ERC20+ERC721    & 18 (2.76\%)  \\
                            & ERC20+ERC1155   & 3 (0.46\%)   \\
                            & ERC721+ERC1155  & 1 (0.15\%)   \\
                            & at least two protocols      & 22 (3.38\%)  \\
                            & three protocols          & 0 (0.00\%)   \\
    \bottomrule
  \end{tabular}
  }
\end{table}
It can be observed that among the ERC20+ERC721 (accounting for 2.76\% of the total) and ERC20+ERC1155 (accounting for 0.46\% of the total) categories, the mixed graph has a higher proportion than the normal graph. In other words, when NFT phishing occurs involving ERC721 or ERC1155 tokens, it is more likely to be accompanied by ERC20 transactions compared to regular accounts. This indicates that NFT phishing frequently entails additional token operations based on the ERC20 standard following the transfer of NFT ownership.

\begin{tcolorbox}[boxrule=1pt,boxsep=1pt,left=3pt,right=3pt,top=3pt,bottom=3pt]
\textbf{Answer to RQ2.}
NFT phishing tends to combine the use of the ERC20 standard with two other ERC standards protocols.
\end{tcolorbox}

\subsection{How does NFT phishing behave?}
\label{subsec:Interaction_Pattern}
To explore the behavior of NFT phishing, we conducted a more detailed analysis of the data of phishing addresses and normal addresses in the mixed graph.

As shown in Table \ref{tab:edge_types}, in the mixed graph, the number of transactions between normal accounts reaches 435, accounting for 66.82\% of the total. The interactions between phishing accounts are only 4 times, indicating that there is relatively little cooperation among these phishing accounts. They are mainly targeting normal addresses. A total of 14.59\% of the transactions are initiated from regular accounts to phishing accounts.

\begin{table}[htbp]
\centering
\caption{Amount and Ratio of Different Types in Mixed Graph}
\label{tab:edge_types}
\begin{tabular}{lcc}
\toprule
\textbf{Types} & \textbf{Amount} & \textbf{Ratio} \\
\midrule
normal$\rightarrow$normal & 435 & 66.82\% \\
phishing$\rightarrow$normal & 117 & 17.97\% \\
normal$\rightarrow$phishing & 95 & 14.59\% \\
phishing$\rightarrow$phishing & 4 & 0.61\% \\
\midrule
\textbf{Total} & 651 & 100.00\% \\
\bottomrule
\end{tabular}
\end{table}

Therefore, by analyzing the interaction information of the mixed graph, we can identify the following features. The mixed graph is mainly composed of dense normal transactions, with the phishing addresses conducting multiple penetrations of specific addresses, and there is relatively little interaction among the phishing addresses.

Furthermore, to conduct a more in-depth analysis of the flow relationships, we count crucial nodes in the mixed graph (i.e., Convergence nodes, Distribution nodes, and Bidirectional Hub).

\begin{table}[htbp]
\centering
\caption{Roles Statistics Analysis of Token Flow}
\label{tab:fund_flow_roles}
\begin{tabular}{lc}
\toprule
\textbf{Role Types} & \textbf{Amount (Ratio)} \\
\midrule
Convergence Nodes & 62 (1.57\%) \\
Distribution Nodes & 80 (2.03\%) \\
Bidirectional Hubs & 553 (14.05\%) \\
\bottomrule
\end{tabular}
\end{table}

Convergence nodes refer to the nodes with high in-degree and low out-degree characteristics. They receive a large amount of funds but rarely have any outflows. 
Distribution nodes are nodes with low in-degree and high out-degree. They transfer a large amount of data outward but rarely receive returns. Bidirectional Hubs refer to nodes with high in-degree and out-degree. Such nodes are associated with the majority of transactions in the network and are the key nodes in the entire network. As shown in the statistics in Table \ref{tab:fund_flow_roles}, there are 1.57\% convergence nodes, 2.03\% distribution nodes, and 14.05\% bidirectional hubs. It indicates that the number of nodes in the network that are highly related to the token flows only constitutes a small portion.

\begin{tcolorbox}[boxrule=1pt,boxsep=1pt,left=3pt,right=3pt,top=3pt,bottom=3pt]
\textbf{Answer to RQ3.}
NFT phishers frequently target particular addresses, and interactions between phishing accounts occur with lower frequency.
\end{tcolorbox} 




\section{Conclusion}
\label{sec:conclusion}
This study aims to fill the gap in the existing literature regarding the in-depth exploration of the on-chain transaction system of NFT phishing scams. By constructing a transaction graph and combining large-scale on-chain data, it reveals the distribution patterns, transaction features, and interaction patterns of NFT phishing scams, to enhance the understanding of such fraudulent behaviors. Through the collection and standardization processing of NFT phishing-related data from multiple platforms, the study ultimately clarified the core behavioral patterns and key features of NFT phishing scams. 
The main results show that although NFT phishing accounts account for a very small proportion of the total accounts, their participation ratio in the transaction scenarios is significantly higher than their account proportion. Moreover, in large-scale transaction networks, the interaction probability with normal accounts has significantly increased. The phishers tend to interact with victims through multi-protocol mixed transactions. At the same time, they pay more attention to carrying out scams against specific normal accounts rather than collaborating among phishing accounts.



\bibliographystyle{IEEEtran}
\bibliography{ref}



\section{Biography Section}
\vspace{-3em}
\begin{IEEEbiography}[{\includegraphics[width=1in,height=1.25in,clip,keepaspectratio]{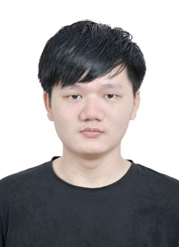}}]{Wenkai Li}
Wenkai Li is currently pursuing a doctor's degree in the School of Cyberspace Security at Hainan University, China. His research lies in smart contract security and malicious behavior analysis, focusing on enhancing blockchain security through software and data analytics. He is also exploring the integration of AI, including GNN and LLM.
\end{IEEEbiography}
\vspace{-3em}
\begin{IEEEbiography}[{\includegraphics[width=1in,height=1.25in,clip,keepaspectratio]{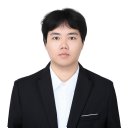}}]{Zongwei Li}
Zongwei Li is currently a Ph.D. student in the School of Cyberspace Security at Hainan University, China. His research interests lie in blockchain security, smart contract vulnerability detection, and the application of AI4SE.
\end{IEEEbiography}
\vspace{-3em}
\begin{IEEEbiography}[{\includegraphics[width=1in,height=1.25in,clip,keepaspectratio]{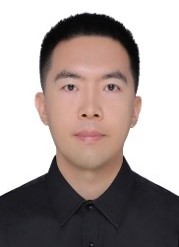}}]{Xiaoqi Li}
Xiaoqi Li is an associate professor at Hainan University. Previously, he was a researcher at the Hong Kong Polytechnic University. He received his Ph.D. in Computer Science from Hong Kong Polytechnic University, MSc in Information Security from the Chinese Academy of Sciences, and BSc in Information Security from Central South University. His current research interests include Blockchain/Mobile/System Security and Privacy, Ethereum/Smart Contract, Software Engineering, and Static/Dynamic Program Analysis. He received best paper awards from INFOCOM'18, ISPEC'17, CCF'18, and an outstanding reviewer award from FGCS'17.
\end{IEEEbiography}
\vspace{-3em}
\begin{IEEEbiography}[{\includegraphics[width=1in,height=1.25in,clip,keepaspectratio]{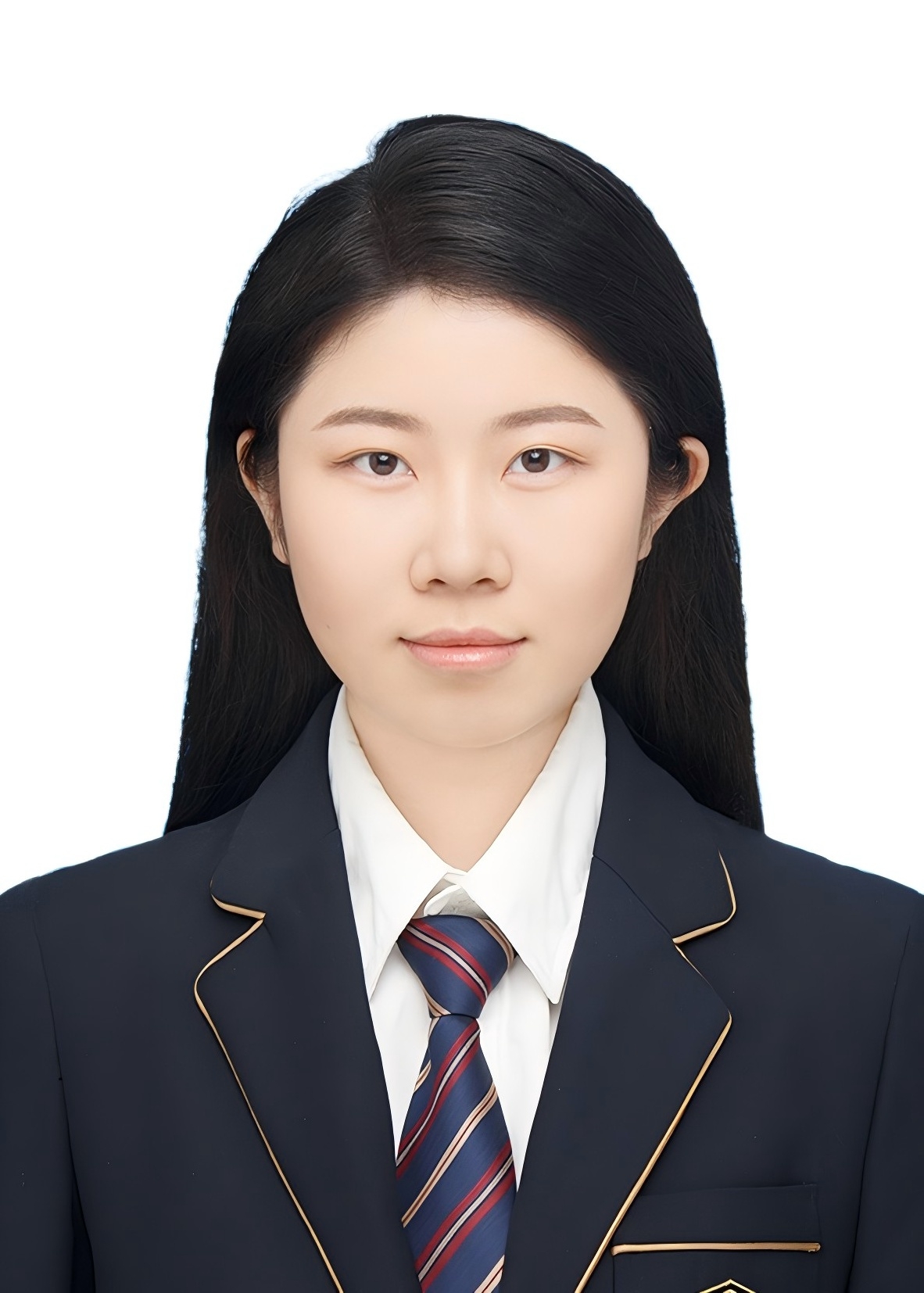}}]{Chunyi Zhang}
Chunyi Zhang is currently pursuing a master's degree in the School of Cyberspace Security at Hainan University, China. Her research interests lie in blockchain security and the application of Large Language Models in cybersecurity.
\end{IEEEbiography}
\vspace{-1em}
\begin{IEEEbiography}[{\includegraphics[width=1in,height=1.25in,clip,keepaspectratio]{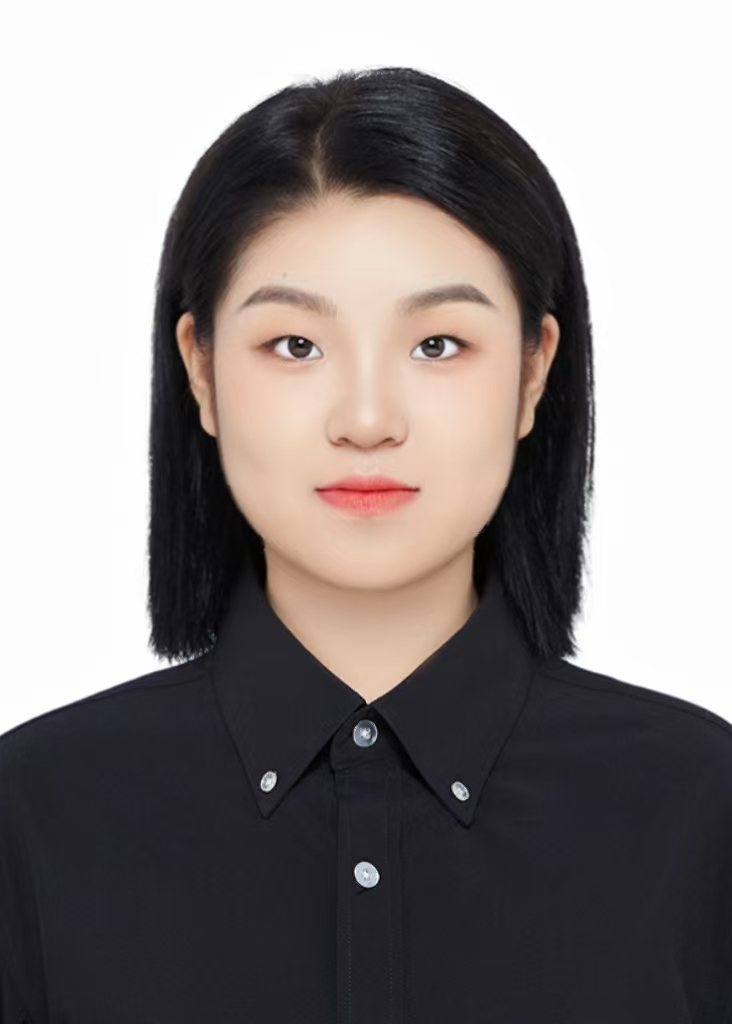}}]{Xiaoyan Zhang}
Xiaoyan Zhang is currently pursuing a master's degree in the School of Cyberspace Security at Hainan University, China. Her current research interests include Blockchain Security and Large Language Models.
\end{IEEEbiography}
\vspace{-1em}
\begin{IEEEbiography}[{\includegraphics[width=1in,height=1.25in,clip,keepaspectratio]{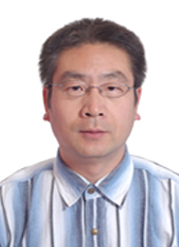}}]{Yuqing Zhang}
Yuqing Zhang is the Director of the Chinese National Computer Network Intrusion Prevention Center, Deputy Director of the Chinese National Engineering Laboratory of Computer Virus Prevention Technology, Vice Dean of the School of Computer and Control Engineering at the Chinese Academy of Sciences, and Professor at Hainan University. He received his Ph.D. from Xi'an University of Electronic Science and Technology. He has presented over 100 papers and 7 national/industry standards. His research interests include Network Attacks and Prevention, Security Vulnerability Mining and Exploitation, IoT System Security, AI Security, Data Security, and Privacy Protection.
\end{IEEEbiography}
\vfill

\end{document}